\documentclass[10pt,letterpaper]{article}
\usepackage{opex3}
\usepackage{color}
\usepackage{threeparttable}
\usepackage[draft]{hyperref}
\usepackage{cite}
\usepackage{amsmath}
\begin{document}

\title{The wave impedance of an atomically thin crystal}

\author{Michele Merano}

\address{Dipartimento di Fisica e Astronomia G. Galilei, Universit$\grave{a}$ degli studi di Padova, via Marzolo 8, 35131 Padova, Italy}

\email{michele.merano@unipd.it} 



\begin{abstract}
I propose an expression for the electromagnetic wave impedance of a two-dimensional atomic crystal, and I deduce the Fresnel coefficients in terms of this quantity. It is widely known that a two-dimensional crystal can absorb light, if its conductivity is different from zero. It is less emphasized that they can also store a certain amount of electromagnetic energy. The concept of impedance is useful to quantify this point.
\end{abstract}




Graphene, the first material consisting of a single plane of atoms, was created in 2004 \cite{Novoselov2004}. Boron-nitride and transition metal dichalcogenides soon followed \cite{Novoselov2005}. These materials can be conductors \cite{Ando2002}, semiconductors \cite{Heinz2010} or insulators \cite{Blake2011} and they have extraordinary optical properties. The fine structure constant defines the optical transparency of graphene \cite{Nair2008}, atomically thin transition metal dichalcogenides are direct band semincoductors \cite{Heinz2010}.

Also the reflection law, the most fundamental phenomenon of light matter interaction, is special in these materials. In a recent paper \cite{Merano15} it was shown that for light, a single plane of atoms has no thickness. It appears as a real two dimensional system. In general the experiments, like optical contrast \cite{Blake2007, Kis11, Sandhu13, Rhee12, Hansen11, Rabe10, Dai15, Nolte09, Rubio10} or ellipsometry \cite{Kravets2010, Heinz2014, Diebold10, Weiss10, Gajic15, Sanden10}, involving reflection \cite{Pasupathy13, Tian13}, transmission and absorption of light on a two-dimensional (2D) crystal, have been modelled by treating it as an homogeneous material with a certain thickness. Indeed when deposited on a substrate, the most common situation for experiments, atomic force microscopy tips can measure the thickness of these materials \cite{Novoselov2004, Novoselov2005}. But, when considering optical experiments, a model based on thickness fails to explain the overall experiments on light matter interaction (for instance absorption) \cite{Merano15}. Instead the Fresnel coefficients provided in \cite{Merano15} were able to give a complete and convincing description of all the experimental observations.

Equations furnishing the Fresnel coefficients were obtained by treating the 2D crystal as a boundary and by imposing the right boundary conditions.  When considering the reflection in between two bulk materials, the boundary conditions are the analog of the Kirchhoff's laws for electric circuits and transmission-lines \cite{Adler}. From this analogy it is possible to introduce characteristic wave impedances for treating the reflection and transmission coefficients for plane waves.  Here I show that this analogy can be extended to define a wave impedance of an atomically thin 2D crystal, and I deduce the reflection and the transmission coefficients in terms of this quantity. Next I show how impedance is related to the average density of the dissipated electromagnetic power and the average density of electric energy stored in the 2D crystal.
\begin{figure}[h]
\centering\includegraphics[width=9cm]{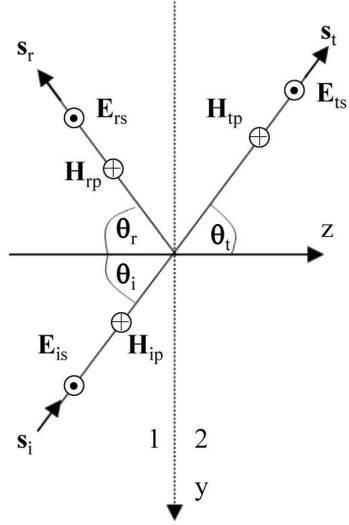}
\caption{Electromagnetic plane wave reflected and transmitted by an interface between two bulk materials separated by an atomically thin 2D crystal. The electric (magnetic) field for $s$, ($p$) polarization is shown.}
\end{figure}

Figure 1 shows a plane wave at oblique incidence on an interface in between two bulk materials separated by a 2D crystal flake. The crystal is assumed to be non magnetic i.e. with a surface magnetic susceptibility $\chi _{m}=0$. The boundary conditions for this case were derived in \cite{Merano15} For $s$ (left column) and $p$ (right column) polarization we have:
\begin{equation}
\begin{aligned}[c]
E_{xi}+E_{xr}&=E_{xt}; \\
E_{xi}+E_{xr}&=\frac{i\omega P_{x}}{i\omega \epsilon_{0}\chi}; \\
E_{xi}+E_{xr}&=\frac{j_{x}}{\sigma}; \\
H_{yi}-H_{yr}&=H_{yt}+i\omega P_{x}+J_{x}; \\
\end{aligned}
\qquad \qquad \qquad
\begin{aligned}[c]
E_{yi}-E_{yr}&=E_{yt} \\
E_{yi}-E_{yr}&=\frac{i\omega P_{y}}{i\omega \epsilon_{0}\chi} \nonumber  \\
E_{yi}-E_{yr}&= \frac{j_{y}}{\sigma} \nonumber \\
H_{xi}+H_{xr}&=H_{xt}+i\omega P_{y}+J{y} \nonumber
\end{aligned}
\end{equation}
where $E_{x}$, $E_{y}$, $H_{x}$, $H_{y}$ are the components of the electric and magnetic fields, $P_{x}$, $P_{y}$, $J_{x}$, $J_{y}$ are the components of the density of polarization and of the density of the conduction current, $\chi$, $\sigma$, $\epsilon_{0}$ are the surface electric susceptibility, the surface conductivity and the vacuum permittivity, and $\omega$ is the angular frequency of the incident light \cite{Merano15}. The electric and magnetic field in bulk media 1 and 2 are connected by:
\begin{equation}
\eta_{1}\vec{\textbf{\emph{H}}}_{i(r)}=\hat{s}_{i(r)}\times\vec{\textbf{\emph{E}}}_{i(r)}; \quad \eta_{2}\vec{\textbf{\emph{H}}}_{t}=\hat{s}_{t}\times\vec{\textbf{\emph{E}}}_{t}
\end{equation}
where $\eta_{1}=\eta /n_{1}$, $\eta_{2}=\eta /n_{2}$, $\eta$ is the impedance of vacuum and $n_{1}$, $n_{2}$ are the refractive indexes of the 2 bulk media.
\begin{figure}[h]
\centering\includegraphics[width=9cm]{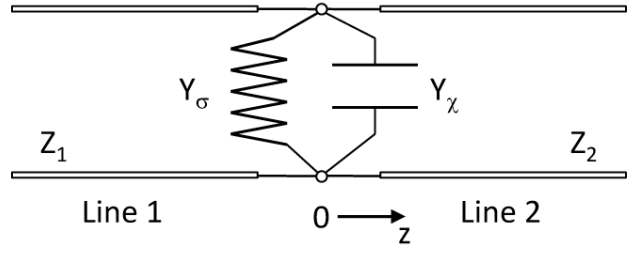}
\caption{Transmission line analogy for the case in Fig. 1. $Z_{1}$ and $Z_{2}$ are the wave impedances in the two media; they are different for the $s$ and $p$ polarization (see text). The atomically thin 2D crystal is analogous to a lumped-circuit load.}
\end{figure}

The analogy with the electric circuits runs as follow(see Fig.2): the two bulk materials are transmission lines with impedance given by $Z_{1s}=\eta_{1}/\cos \theta_{i}$ and $Z_{2s}=\eta_{2}/\cos \theta_{i}$ for $s$ polarization, and $Z_{1p}=\eta_{1}\cos \theta_{i}$ and $Z_{2p}=\eta_{2}\cos \theta_{i}$ for $p$ polarization \cite{Adler}, the 2D crystal is the analog to a lumped-circuit element in between the two with admittance $Y =\sigma + i\omega \epsilon_{0} \chi = Y_{\sigma} +Y_{\chi}$. In practice it is a load consisting of a resistor ($Y_{\sigma}$) and a capacitor ($Y_{\chi}$) placed in parallel in between the 2 transmission lines. This can be easily seen, if we consider the electric and magnetic fields as the analogous of the voltage and the current in the transmission lines, and $J_{tot}=J_{x (y)}+i\omega P_{x (y)}$ as the current in the load for $s$ ($p$) polarization. By simple algebra the boundary conditions for $s$ (left column) and $p$ (right column) polarization become:
\begin{equation}
\begin{aligned}[c]
&E_{xi}+E_{xr}=E_{xt}; \\
&Y_{\chi}(E_{xi}+E_{xr})=i\omega P_{x}; \\
&Y_{\sigma}(E_{xi}+E_{xr})=j_{x}; \\
&\frac{E_{xi}+E_{xr}}{Z_{1s}}=\frac{E_{xt}}{Z_{2s}}+i\omega P_{x}+J_{x}; \\
\end{aligned}
\qquad \qquad \qquad
\begin{aligned}[c]
&Z_{1p}\cdot (H_{xi}-H_{xr})=Z_{2p}\cdot H_{xt}  \\
&Y_{\chi}Z_{1p}\cdot (H_{xi}-H_{xr})=i\omega P_{y} \nonumber  \\
&Y_{\sigma}Z_{1p}\cdot (H_{xi}-H_{xr})= j_{y} \nonumber \\
&H_{xi}+H_{xr}=H_{xt}+i\omega P_{y} +J_{y} \nonumber
\end{aligned}
\end{equation}
The electric field is continuous across the boundary (z=0 in Fig. 2), it fixes the current densities due to the bound and free charges via the admittances $Y_{\sigma}$, $Y_{\chi}$, while the magnetic field jumps of a quantity fixed by $J_{tot}$. The incident electromagnetic field generates a reflected and a transmitted field to adapt to the junction plus the load.

The reflection and the transmission coefficients \cite{Merano15} given by $r_{s}=E_{r}/E_{i}$, $t_{s}=E_{t}/E_{i}$ and $r_{p}=H_{r}/H_{i}$, $t_{p}=H_{t}/H_{i}$ are:
\begin{equation}
r_{s}=\frac{Z_{s2}-Z_{s1}-Z_{s1}Z_{s2}Y}{Z_{s2}+Z_{s1}+Z_{s1}Z_{s2}Y}; \quad r_{p}=\frac{Z_{p1}-Z_{p2}+Z_{p1}Z_{p2}Y}{Z_{p1}+Z_{p2}+Z_{p1}Z_{p2}Y}
\end{equation}
and $t_{s}=r_{s}+1$, $t_{p}=(1-r_{p})Z_{1p}/Z_{2p}$. For $Z_{s1}=Z_{s2}$, $Z_{p1}=Z_{p2}$ we retrieve the formulas of a free standing layer, $\sigma =0$ fixes the case for insulators.

I show now that the real part of the impedance of an atomically thin 2D crystal is twice the average power dissipated in the crystal for an amp$\rm \grave{e}$re/m of $J_{tot}$, whereas the reactance is proportional to the average density of electric energy stored due to the same $J_{tot}$. I start by computing the average density of the electromagnetic power $\Pi$ transformed into heat by the 2D crystal and the reactive power density Q \cite{Fano}. I consider $s$ polarization:
\begin{equation}
\Pi+iQ=\frac{1}{2}{E}_{xt}{J^*}_{tot}=\frac{1}{2}(Y_{\sigma}-Y_{\chi})E_{xt}E^*_{xt}=\frac{1}{2}(\sigma-i\omega \epsilon_{0} \chi) \mid E^2_{xt} \mid
\end{equation}
I note that I can write $Q$ as:
\begin{equation}
Q=-\omega \epsilon_{0} \chi\mid E^2_{xt} \mid=-\omega P_{x}E^*_{xt}=-\omega W_{e}
\end{equation}
where $W_{e}$ is the average density of electric energy per unit surface. In strict analogy with electric circuits \cite{Adler} the impedance $Z = 1/Y$ is connected to the average density power by:
\begin{equation}
\frac{2\cdot (\Pi+iQ)}{\mid J^2_{tot}\mid }=\frac{\sigma-i\omega \epsilon_{0} \chi}{\mid (\sigma+i\omega \epsilon_{0} \chi)^2\mid }=Z
\end{equation}
For $p$ polarization the equations are exactly the same, it is enough to replace $E_{xt}$ with $E_{yt}$. If we put $\sigma = 0$, we obtain the case for an insulator, for which $\Pi=0$ \cite{Blake2011} but $Q\neq0$, so an insulating 2D crystal does not dissipate any power but it keeps storing an electric energy.

I have introduced the concept of the wave impedance of a 2D atomically thin crystal. It is the ratio of the total electric field in the material divided by the total surface density current (due to bound and free charges). From dimensional considerations this definition is specific of a 2D material. I showed that atomically thin crystals act as the analogous of lumped-circuit elements for transmission lines. Usually these materials have been described as homogeneous media with thickness $h$. In transmission lines analogy, these correspond to a line of length $h$ \cite{Adler}, where multiple reflections occur, whereas in a lumped circuit element, and hence in a 2D atomic crystal, multiple reflections do not occur. The concept of impedance proposed here shows once more how extraordinary is the optical behavior of a material composed of a single plane of atoms.

\end{document}